\documentstyle[12pt, epsfig, aaspp]{article}

 1

\def\kms{\,{\rm {km\, s^{-1}}}}

\def\Gyr{{\,\rm Gyr}}
\def\etal{{et al. }}
\def\HH{${\rm {H_2}}\,\,$}
\def\erg{{\rm erg}}

\def\cm{{\rm cm}}

\def\gs{\mathrel{\raise1.16pt\hbox{$>$}\kern-7.0pt
\lower3.06pt\hbox{{$\scriptstyle \sim$}}}}
\def\ls{\mathrel{\raise1.16pt\hbox{$<$}\kern-7.0pt
\lower3.06pt\hbox{{$\scriptstyle \sim$}}}}
\def\gtsima{$\; \buildrel > \over \sim \;$}
\def\ltsima{$\; \buildrel < \over \sim \;$}
\def\prosima{$\; \buildrel \propto \over \sim \;$}
\def\gsim{\lower.5ex\hbox{\gtsima}}
\def\lsim{\lower.5ex\hbox{\ltsima}}
\def\simgt{\lower.5ex\hbox{\gtsima}}
\def\simlt{\lower.5ex\hbox{\ltsima}}
\def\simpr{\lower.5ex\hbox{\prosima}}

\def\pp{\noindent\parshape 2 0truecm 17truecm 2truecm 15truecm}
\def\rf#1;#2;#3;#4 {\par\pp#1, #2, #3, #4. \par}

\def\pr{\ref@jnl{Phys.Rev}}

\newcommand{\href}[2]{{\bf #2} (\texttt{#1})}

\def\ssubsection#1 {\subsection{\sl #1}}

\def\beq#1{\begin{equation}\label{#1}}
\def\eeq{\end{equation}}
\def\beqa#1{\begin{eqnarray}\label{#1}}
\def\eeqa{\end{eqnarray}}
\def\eq#1{equation~(\ref{#1})}

\def\tento#1{\times 10^{#1}}

\def\Ms{\ M_{\odot}}

\def\K{{\rm \ K}}
\def\s{{\rm \ s}}

\def\ergs{{\rm \ erg}}
\def\erg{{\rm \ erg}}
\def\cm{{\rm \ cm}}

\def\kpc{{\rm \ kpc}}

\def\yrs{{\rm \ years}}

\def\HH{H$_2$ }
\def\H2p{H$_2^+$ }
\def\HHp{H$_2^+$ }

\def\Hm{H$^-$ }

\def\mH2p{H_2^+}

\begin{document}

\title{\Large\bf First Structure Formation: \\
  I. Primordial Star Forming Regions in hierarchical models.}

\author{Tom Abel$^{1,2}$,  Peter Anninos$^1$, Michael L. Norman$^1$,
       \& Yu Zhang$^1$  \\
\texttt{abel@mpa-garching.mpg.de, panninos@ncsa.uiuc.edu, \\
norman@ncsa.uiuc.edu, yzhang@ncsa.uiuc.edu}}
\vspace{1cm}
\affil{\vspace{8cm}
\it \flushleft $^1$Laboratory for Computational Astrophysics \\
National Center for Supercomputing Applications \\
University of Illinois at Urbana--Champaign \\
405 N. Mathews Ave., Urbana, IL 61801\\
\ \\
$^2$Max Planck Institut f\"ur Astrophysik \\
Karl Schwarzschildstr. 1, Postfach 1523, \\
85740 Garching, Germany\\
\vspace{5cm}}

\begin{abstract}

  We investigate the possibility of very early formation of primordial
  star clusters from high--$\sigma$ perturbations in cold dark matter
  dominated structure formation scenarios. For this we have developed
  a powerful 2--level hierarchical cosmological code with a realistic
  and robust treatment of multi--species primordial gas chemistry,
  paying special attention to the formation and destruction of \HH
  molecules, non--equilibrium ionization, and cooling processes. We
  performed 3--D simulations at small scales and at high redshifts and
  find that, analogous to simulations of large scale structure, a
  complex system of filaments, sheets, and spherical knots at the
  intersections of filaments form. On the mass scales covered by our
  simulations ($5\tento{5} - 1\tento{9}\Ms$) that collapse at
  redshifts $z>25$, we find that only at the spherical knots can
  enough \HH be formed ($n_{H_2}/n_H \gsim 5\tento{-4}$) to cool the
  gas appreciably. Quantities such as the time dependence of the
  formation of \HH molecules, the final \HH fraction, and central
  densities from the simulations are compared to the theoretical
  predictions of Abel (1995) and Tegmark \etal (1997) and found to
  agree remarkably well.  Comparing the 3--D results to an isobaric
  collapse model we further discuss the possible implications of the
  extensive merging of small structure that is inherent in
  hierarchical models. Typically only 5--8\% percent  of the total
  baryonic mass in the collapsing structures is found to cool
  significanlty. Assuming the Padoan (1995) model for star formation
  our results would predict the first stellar systems to be as small
  as $\sim 30\Ms$. Some implications for primordial globular cluster
  formation scenarios are also discussed.

\end{abstract}

\section{INTRODUCTION}
\label{sec:introduction}

Our primary interest is to study and identify the physics responsible
for the formation of the first star in the universe. If the models put
forward in previous investigations are correct, this is a very well
defined problem because, for a given cosmogony, the thermodynamic
properties, baryon content, and chemical composition are specified for
the entire universe at high redshifts, at least in a statistical
sense.  Furthermore, the dominant physical laws are readily identified
with the general theory of relativity describing the evolution of
the background space--time geometry
and particle geodesics, the Euler equations governing the
motion of the baryonic fluid(s) in an expanding universe, and the
primordial kinetic rate equations determining the chemical processes.
Hence, at least in principal, given accurate numerical methods, one
can directly simulate the formation of the first star in the universe.
In practice, however, it turns out that the dynamical range required
is beyond what current state--of--the--art numerical methods and
super--computers can achieve. Nevertheless, it is worth stressing that
the study of primordial star formation can yield stringent theoretical
constraints on current cosmogonies.  In fact it can even discard
theories of structure formation if it is found that star formation is
predicted to occur too late to explain the observed high redshift
universe, such as the metal absorption lines present
in quasar spectra at high redshifts.

Models for structure formation are based on the growth of small
primordial density fluctuations by gravitational instabilities on a
homogeneously expanding background universe.  In all scenarios, the
minimum mass scales to separate from the Hubble flow and begin
collapsing are the smallest ones which survived damping until after
the recombination epoch.  Depending on the nature of the dark matter
and whether the primordial fluctuations were adiabatic or isothermal,
this mass scale could be as small as one solar mass (very heavy cold
dark matter (CDM) particles) or as high as $\gsim 10^{13}M_\odot$ (HDM
scenarios).  In CDM cosmogonies, the fluctuation spectrum at small
wavelengths has only a logarithmic dependence for mass scales smaller
than $\sim 10^8M_\odot$, which indicates that all small scale
fluctuations in this model collapse nearly simultaneously in time.
This leads to very complex dynamics during the formation of these
structures.  Furthermore, the cooling in these fluctuations is
dominated by the rotational/vibrational modes of hydrogen molecules
that were able to form using the non--equilibrium abundance of free
electrons left over from recombination and the ones produced by strong
shock waves as catalysts (Saslaw \& Zipoy, 1967; Couchman \& Rees,
1986; Tegmark \etal 1997).

Couchman \& Rees (1986) have argued that the first structures to
collapse might have substantial influence on the subsequent thermal
evolution of the intergalactic medium due to the radiation they emit
as well as supernova driven winds.  Haiman \etal (1997) have argued
that the fragmentation of the first collapsed objects into stars can
lead to the photoionization of the IGM and raise its temperature to
$\sim 10^4$ K, and the Jeans mass to $M_J \gg 10^5~M_\odot$.  Such an
ionizing background can have a negative feedback effect on the
molecular hydrogen fraction since \HH can be destroyed by photons in
its Lyman--Werner Bands which are below the Lyman limit. Accounting for
self--shielding, Haiman \etal (1997) demonstrate that \HH molecules
inside virialized clouds are easily destroyed before the universe
fully reionizes.  The appearance of the first UV sources will thus
suppress the \HH abundance in primordial structures and prevent the
post--virialization collapse.  \HH cooling is then also suppressed,
even in massive objects, before the UV radiation photoionizes the
entire IGM.  The only clouds that are able to fragment into new stars
have $T_{vir}=10^4$ K and nonlinear mass scales above the new Jeans
mass $M > 10^8 (1+z_{vir}/10 )^{-3/2}~M_\odot$.  These conclusions
will hold for structures that are collapsing after a significant UV
background flux has been established.  However, if structures collapse
and cool to number densities $\gsim 1 \cm^{-3}$ with neutral hydrogen
column densities $\gsim 10^{18}\cm^{-2}$ before the onset of the UV
flux, they are able to survive quite strong UV fluxes ($\gsim 10^{-22}
\erg \cm^{-2} \s^{-1}$) (Haiman \etal 1996a; Abel \& Mo 1997).  Hence
to know what structures will really be destroyed by the UV background,
one first has to answer a crucial question: {\em How long does it take
to form the first stars in the universe?}.

The pioneering studies on primordial star formation ({\it i.e.}
Hutchins 1976; Palla \etal 1983) explored various aspects of the
microphysical processes during protostellar collapse, the minimum
attainable Jeans mass, and the accretion stage of primordial star
formation.  However, the time scale for the formation of a primordial
star could not yet be determined due to the complexity and
interactions of the hydro--dynamical, chemical, and radiative
processes.  The typically high computational requirements to solve the
chemical rate equations alone have forced many previous studies to use
the steady state shock assumption, or a spherical collapse model which
formulate the problem without any spatial  dimension.  Only recently
Bodenheimer \etal (1986) and Haiman \etal (1996b) were able to
address the problem with spherically symmetric 1--D hydrodynamical
calculations, including the non--equilibrium chemistry and cooling
processes.

We have been able to develop numerical methods (Anninos, Zhang, Abel,
\& Norman 1997a) and an accurate chemistry and cooling model (Abel,
Anninos, Zhang, \& Norman 1997) that allow us to study the problem in
three dimensions without imposing an equilibrium or other simplifying
assumptions. Recently Gnedin and Ostriker (1997) have also
developed a similar method allowing to model the
non--equilibirum chemistry in multi--dimensional, hydrodynamical
calculations.  Here we use our code to study the origin of the first
structures in the universe formed in the standard cold dark matter
scenario (SCDM).  Although the SCDM model fails to explain
the observed large scale structure, it is still the prime example for
any hierarchical clustering scenario and matches the observations of
small scale systems remarkably well (see Zhang, Anninos \& Norman
1995, and references therein). 

This paper complements the first reported result from this work
presented in Zhang~\etal (1996).  In section~\ref{sec:AModel} we
introduce an analytical description of small scale structure collapse
based on the simplified studies of Abel (1995) and Tegmark \etal
(1997).  Section \ref{sec:methods} discusses our numerical methods and
presents and motivates our simulation parameters.  In
section~\ref{sec:results} we show results from our 3--D investigations
that illustrate the morphology of the first primordial star forming
regions as well as comparing these results with the analytical picture
outlined in section \ref{sec:AModel}.  We find remarkable agreement
with the rough estimates of both Abel (1995) and Tegmark \etal (1997).
We summarize our findings and comment on primordial Globular Cluster
formation scenarios and models of primordial star formation in section
\ref{sec:discussion}.

\section{Analytical Considerations for Small Scale Structure Collapse}
\label{sec:AModel}

\ssubsection{The Density at Collapse}

For the   collapse  of structures   where the initial   gas pressure is
dynamically important, a simple estimate of  the density at collapse is
derived by solving  the equations of hydrostatic  equilibrium (Sunyaev
\&   Zel'dovich  1972;  Tegmark  \etal   1997).   Defining  the virial
temperature $T_{vir}$ through $3 kT_{vir}/2 =  -  m_p \Phi/2$, where
$\Phi$ is the gravitational  potential,  the  density
contrast $\delta$ can be written as
\begin{equation}
  \label{delta}
  \left( 1+ \delta \right) = \frac{\rho}{\rho_B} = \left( 1+
    \frac{6}{5} \frac{T_{vir}}{T_B} \right)^{3/2},
\end{equation}
where $T_B$   and   $\rho_B$ denote  the background   temperature and
density, respectively.  The morphological
characteristics of  the   object described  by  equation \eq{delta}
is determined by the gravitational potential
$\Phi$.  The factor $1/2$  on the  right  hand side of  the
definition  of  $T_{vir}$   is  chosen  to  reproduce  the   canonical
definition of  $3 kT_{vir}/2 = m_p v^2/2  $ for a virialized spherical
system  with velocity dispersion   $v$.  The analysis  of  a spherical
collapse  model for structures  where the initial  gas pressure is not
important gives for  the overdensity  at  virialization $\delta =   18
\pi^2$ (Gunn and Gott 1972).
Hence   \eq{delta} should  only  be applicable as  long as
$\delta   \lsim 18  \pi^2$.   Furthermore  the  assumption of  hydrostatic
equilibrium,  which we  used to derive   \eq{delta},  can only  hold for
structures with infall (virial) velocities that are small compared to the sound
speed of the gas; meaning that no virialization shock will form.  This
is  true for  spherical  objects with  masses $\lsim   100  \Ms h^{-1}
\Omega_0^{-0.5} (1+z_{coll})^{3/2}$, which is in reasonable agreement
with the condition $\delta   \lsim 18  \pi^2$.

The matter temperature of  the background universe decouples thermally
from   the radiation at $z  \sim   200$ and evolves adiabatically  ($T
V^{\gamma -1 } =const.$) afterwards. Hence we can write
\begin{equation}
  \label{backgroundT}
  T_B = 135 \K \left( \frac{1+z}{100} \right)^2,
\end{equation}
with which we find for the density at collapse,
\begin{eqnarray}
  \label{ncollapse}
  n_{coll} = \left\{
      \begin{array}{ll}
        1 \cm^{-3}
        \left(\frac{T_{vir}}{400 \K} \right)^{3/2} \left(
          \frac{\Omega_B h^2}{0.025} \right),  &
        \ \ \ {\rm for \ \ } \frac{T_{vir}}{400 \K} < \left(
          \frac{1+z}{30} \right)^2 \\
        1 \cm^{-3} \left(\frac{1+z}{30} \right)^2 \left(
          \frac{\Omega_B h^2}{0.025} \right), &  \ \ \ {\rm
          otherwise.}
      \end{array}
      \right.
\end{eqnarray}
This suggests that, for the smallest structures, the typical density at
collapse  does  {\em not\/} depend   on the collapsing redshift, and
that the  chemical  and  cooling  behavior for these
structures will therefore  be the same for  any collapse redshift.
However, the
above analysis is  certainly oversimplified in the  sense that it does
not include a detailed treatment of  the dark matter   potential in which  the
baryons  will  fall.

\ssubsection{Molecular Chemistry and Cooling}
\label{sec:H2formation}

Molecular hydrogen can not be destroyed efficiently in primordial gas
at low temperatures ($<$ few $\times 10^3$K) unless there is a
radiation flux higher than $\sim 3 \times 10^{-26} \ergs \cm^{-2}
\s^{-1} $ in the Lyman Werner Bands.  Once self--shielding is
important even higher fluxes would be needed.  Furthermore only two
H$_2$ producing gas phase reactions operate on time--scales less than the
Hubble time: the charge exchange reaction
\begin{equation}\label{reac:H2PH}
\rm
H_2^+ \ \ \ + \ \ \ H \ \ \ \rightarrow \ \ \ H_2 \ \ \ + \ \ \ H^+ ,
\end{equation}
and the dissociative attachment reaction
\begin{equation}\label{reac:HMH}
\rm
H^- \ \ \   + \ \ \ H \ \ \ \rightarrow \ \ \ H_2 \ \ \ + \ \ \ e^-.
\end{equation}
The H$_2$ abundance will thus be determined by the H$_2^+$ and H$^-$
abundances. However, one typically finds that the H$_2^+$ equilibrium
abundance is much lower than that of H$^-$ ($k_9\ll k_7$ in the
notation of Abel \etal 1997a) and additionally that the rate
coefficient of reaction (\ref{reac:H2PH}) is about $2.4$ times less
than reaction (\ref{reac:HMH}).  Hence the H$_2$ formation is
dominated mostly by reaction (\ref{reac:HMH}).

In the absence of an external UV  background at gas temperatures below
6000K, one can integrate  the rate equations for  the free electron and
the molecular hydrogen fraction to find  the time evolution of the \HH
fraction during the  collapse of primordial  gas clouds (see Abel 1995
and Tegmark \etal 1997)
\begin{eqnarray}\label{equ:fH2}
f_{H_2}(t)  &=& f_{H_2}(t=0) + \frac{k_{PA}}{k_{rec}} \ln(x_0
n_H k_{rec} t + 1) \nonumber \\
 & = & f_{H_2}(t=0) + 1.0 \times 10^{-8} T_{vir}^{1.53}\ln(t/t_{rec}+1) ,
\end{eqnarray}
where $k_{PA}$, $k_{rec}$, $t_{rec}$, $x_0$, and $n_H$ are the rate
coefficients of photo--attachment of \Hm and recombination to neutral
hydrogen in $\cm^3\s^{-1}$, the initial recombination timescale (see
\eq{equ:trec}), the initial ionized fraction, and the neutral hydrogen
number density (in cm$^{-3}$), respectively.  The production of \HH
therefore only depends logarithmically on time, and increases most
rapidly within the first few initial recombination timescales so that
it defines an \HH formation timescale,
\begin{equation}\label{equ:trec}
t_{rec} = \frac{1}{n_{H^+} k_{rec}} = 5.0 \times 10^{13}
   \left(\frac{n_H}{100{\rm cm}^{-3}} \right)^{-1} \left(\frac{x}{10^{-4}}
   \right)^{-1} \left( \frac{T}{1000{\rm K}}\right)^{0.65}\ \ {\rm s},
\end{equation}
where $k_{rec} = 1.8 \times 10^{-10} T^{-0.65}\cm^3\s^{-1}$ is our own
fit to the data of Ferland \etal (1992), which is accurate to within a
few percent for temperatures below $10^4$K.
A good estimate for  the typical \HH  fraction is
given by evaluating \eq{equ:fH2} at $t\sim 2 t_{rec}$.
The temperature
dependence enters from the ratio of the recombination and \Hm
formation rates.  A typical \HH fraction of $n_{H2}/n \sim 10^{-3}$ is
produced during the collapse of structures with virial temperatures
greater than $ 10^3 \K$.  For temperatures higher than $6000\K$, the
charge exchange with protons will efficiently destroy H$_2$, and
\eq{equ:fH2} will not be applicable.  However, during the collapse of
clouds with such high virial temperatures, the final \HH fraction will,
nevertheless, be $f_{H_2}(T\sim 6000\K ) \sim 6 \tento{-3}$ (see Abel
\etal 1997a).  Comparing \eq{equ:fH2} with the dynamical timescale
for
a spherical top--hat model, Tegmark \etal (1997) argued for a rule of
thumb, stating: ``A spherical primordial gas cloud will be able to
undergo a stage of free-fall collapse if its \HH fraction becomes
$\gsim 5\tento{-4}$.''

\ssubsection{Isobaric Collapse}
\label{sec:isobaric}

It has been found by numerous authors (see e.g.  Mac Low and Shull,
1986; Shapiro and Kang, 1987; Anninos \& Norman, 1996,
Abel~\etal~1997b) that the postshock gas of a radiative shock
originating in structure formation evolves under constant pressure
($\rho_b \times T \propto p = const$, where $\rho_b$, $T$ and $p$ are
the density, temperature and pressure of the baryonic fluid).
Although there is no reason to assume that isobaric evolution is also
appropriate for spherical collapsing structures it is useful to
compare our simulations to such simplified models.
Constant pressure implies that the
internal energy density of the gas, $e$, also remains constant
(assuming an ideal gas for which $e= p/(\gamma-1)$).  Differentiation
of the isobaric condition yields the following form of the constant
pressure condition
\begin{equation}\label{equ:isoC}
  \frac{\delta \rho_b}{\rho_b} = - \frac{\delta T}{T}.
\end{equation}
The temperature evolution   equation is found   using the energy
conservation    equation   that  neglects  conduction,
viscosity, and kinetic energy terms
\begin{equation}\label{equ:energy}
  \frac{du}{dt} = -\frac{p}{\rho_b^2}
                   \frac{d \rho_b}{dt} + \frac{\Lambda}{\rho_b},
\end{equation}
where  $u  = e/\rho_b$ is   the energy  per  unit mass  and $\Lambda  =
\dot{e}$    the  cooling function  of the   gas.    Inserting the
isobaricity condition (\ref{equ:isoC}) into \eq{equ:energy}, we derive
the following simple ordinary differential equation for the
temperature evolution:
\begin{equation}
  \label{Tdot}
  \frac{\dot{T}}{T} = \frac{\dot{e}(n, T)}
  {\gamma \ e(n, T)} \ \ \ \leftrightarrow
  \ \ \ \dot{T} = \frac{T}{\gamma \ t_{cool}(p_0/T, T)}.
\end{equation}
Here we substituted  the  temperature for the density using  $n=p_0/T$,
where $p_0 =  T_0 n_0$ is the  initial pressure at collapse
and $\gamma$ ($= 5/3$ for an ideal gas)    is the adiabatic index.
The isobaricity condition
($nT = n_{vir}T_{vir}$ throughout the contraction) implies that the
density at the final stages of the isobaric collapse will be $n \sim
n_{vir} T_{vir}/50\K$
since  \HH cooling  becomes  inefficient   at
temperatures   $\lsim 50\K$.

\section{THE SIMULATIONS}
\label{sec:methods}

A numerical code with high spatial  and mass resolution is required to
model both the collapse of sufficiently  large scale structure and the
microphysics of chemical  reactions  and radiative cooling   which are
important on the smaller scales.  We can achieve high dynamical ranges
with the  two--level  hierarchical three--dimensional code  (HERCULES)
that we have developed for cosmology  (Anninos, Norman \& Clarke 1994;
Anninos, Zhang,   Abel  \& Norman 1997).    This code  is  designed to
simulate structure  formation  in an  expanding  dark matter dominated
universe with Newtonian gravity, multi--fluid hydrodynamics, radiative
cooling, non--equilibrium chemistry and external radiation fields.  It
also utilizes nested grid  methods for both the   gas and dark  matter
particles to achieve higher  resolution over a selected sub--domain of
the  coarser  parent grid.    At  the  small  scales  resolved  by our
calculations, it is important to  track the different chemical species
in order to model the chemistry and  radiative cooling accurately. For
this purpose we independently    evolve the following   nine  species:
neutral hydrogen H,   ionized  hydrogen H$^+$, negative hydrogen  ions
H$^-$, hydrogen molecules  H$_2$, ionized hydrogen molecules  H$_2^+$,
neutral helium  He,   singly--ionized helium  He$^+$,  doubly--ionized
helium  He$^{++}$, and  free electrons  $e^-$.   The 28 most  important
chemical rate equations  (including radiation processes) are solved in
non--equilibrium for the abundances of each of the  nine species.  The
rate  coefficients used in the chemistry   model are provided in Abel,
Anninos,  Zhang    \&  Norman  (1997a).  We  have    also implemented a
comprehensive model  for  the radiative cooling  of  the gas (Anninos,
Zhang,   Abel \& Norman   1997) that includes  atomic line excitation,
recombination, collisional   ionization,      free--free  transitions,
molecular line  excitations,  and Compton   scattering of the   cosmic
background radiation (CBR) by electrons.

We apply our code to high redshift pre--galactic structure formation
and evolution, investigating specifically the collapse of the first
high--$\sigma$ bound objects with baryonic masses in the range $10^4 -
10^8 M_\odot$.  Our model background spacetime is a flat
($\Omega_0=1$) cold dark matter dominated universe with Hubble
constant $H_0=50$ km~s$^{-1}$~Mpc$^{-1}$ and baryon fraction
$\Omega_B=0.06$, consistent with the constraints from big--bang
nucleosynthesis (Copi, Schramm \& Turner 1995).  The baryonic matter
is composed of hydrogen and helium in cosmic abundance with a hydrogen
mass fraction of 76\% and ratio of specific heats $\gamma=5/3$.  The
data is initialized at redshift $z=100$ with matter perturbations
derived from the Harrison--Zel'dovich power spectrum modulated with a
transfer function appropriate for CDM adiabatic fluctuations and
normalized to the cluster scale $\sigma_{8h^{-1}}=0.7$.
Bertschinger's (1994) constrained realization procedure is used to
construct 3 and 4$\sigma$ fluctuations at the box center over a region
of 1/5 the box length.

The  chemical abundances  are initialized  according  to the estimates
provided by  Anninos \&  Norman (1996). For  the primary  hydrogen and
helium species, we use  the  residual ionization estimate of   Peebles
(1968, 1993):
\begin{equation}
\label{resion}
\frac{n_{H^+}}{n_H} = 2.4 \times 10^{-4}~ \Omega_0^{1/2}
   \frac{0.05}{h\Omega_B},
\end{equation}
and assume negligible amounts of $n_{He^+}/n_H$ and $n_{He^{++}}/n_H$
($\lesssim 10^{-14}$). The concentrations of the intermediaries
H$^-$ and H$_2^+$, and hydrogen molecules are given by
\begin{eqnarray}
\frac{n_{H^-}}{n_H}
                 & = & 2\times 10^{-9}~T_K^{0.88}~\frac{n_{H^+}}{n_H}~, \\
\frac{n_{H_2^+}}{n_H}
                 & = & 3\times 10^{-14}~T_K^{1.8}~\frac{n_{H^+}}{n_H}~, \\
\frac{n_{H_2}}{n_H} & = & 2\times 10^{-20}
                      \frac{f_H \Omega_0^{3/2}}{h\Omega_B} (1+z_o)^{5.1}~,
\end{eqnarray}
where $T_K$ is the gas  temperature in degrees  Kelvin, and is set  at
the starting redshift by assuming an adiabatic expansion from $z=200$,
the  approximate  redshift at  which matter/radiation interactions are
negligible (see \eq{backgroundT}).   Also, $f_H$ is the hydrogen  mass
fraction and $z_o =  300$ is an  estimate for the
largest   redshift at which H$_2^+$   can  form efficiently to produce
hydrogen  molecules  without   being photodissociated by   the  cosmic
background  radiation.  For our  model    parameters ($f_H =    0.76$,
$\Omega_0=1$, $\Omega_B=0.06$  and  $h=0.5$), we find  $n_{H_2}/n_H  =
2.9\times 10^{-6}$.  The neutral   hydrogen and helium, and   electron
densities are then set by the following conservation equations:
\begin{equation}
  \rho_{He} + \rho_{He^+} + \rho_{He^{++}} = \rho_b~(1-f_H)~,
\end{equation}
\begin{equation}
  \rho_H + \rho_{H^+} + \rho_{H^-} + \rho_{H_2^+} + \rho_{H_2}
      = \rho_b~f_H~,
\end{equation}
\begin{equation}
  \rho_{H^+} - \rho_{H^-} + \frac{1}{2}\rho_{H_2^+} +
      \frac{1}{4}\rho_{He^+} + \frac{1}{2}\rho_{He^{++}} = m_H~n_e~,
\end{equation}
where $\rho_i$ are the densities of the $i$th species, $\rho_b$ the
total baryonic density, $n_e$ the number density of free electrons,
and $m_H$ the proton mass.

In this  paper we  present   results from  six  different  nested grid
calculations.  We investigate the effect of increasing the large scale
power by varying the parent box size over $L=$ 128,  512 and 1024 kpc,
and set  up initial  data  for  the  collapse  of both  3$\sigma$  and
4$\sigma$ fluctuations.   The different box  sizes act to parameterize
the mass of  the collapsing structures.   The  data from the
sub grid (top grid) simulation of a $4\sigma$  peak within a box of size
512kpc  is referred to as $4\sigma 512$S ($4\sigma  512$T).
For a  fixed mass scale,
$\sigma$ effectively defines the   redshift  at which the   structures
begin to collapse.  A  box dimension of 1024 kpc  is about the maximum
we can set before   losing the resolution  in the  sub grid needed  to
resolve  the H$_2$  chemistry and  cooling.   We use $128^3$ cells  to
cover both the  top and sub grids which,  for the refinement factor of
four  used in  all our  simulations, results in   an effective $512^3$
resolution over the sub grid domain.  The spatial and mass resolutions
of the parent and child grids are  shown in Table~\ref{tab:resolution}
for each of the calculations.
\begin{table}[htb]
\centerline{\begin{tabular}{|c|c|c|c|c|c|c|c|}
\hline
\hline
Run & Grid & $L$ (kpc) & $\Delta x$ (kpc) & $M_{DM}$ ($M_\odot$) & $M_{B}$
($M_\odot$)
&
$\Delta M_{DM}$ ($M_\odot$) & $\Delta M_{B}$ ($M_\odot$)  \\
\hline
3(4)$\sigma$1024T & top & 1024 & 8    & $7.5\times10^{10}$ & $4.5\times10^{9}$
                  & $3.6\times10^{ 4}$ & $2.1\times10^{3}$   \\
3(4)$\sigma$512T  &  top &  512 & 4    & $9.3\times10^{ 9}$ & $5.6\times10^{8}$
                  & $4.4\times10^{ 3}$ & $2.7\times10^{2}$   \\
3(4)$\sigma$128T  &  top &  128 & 1    & $1.5\times10^{ 8}$ & $8.8\times10^{6}$
                  & $6.9\times10^{ 1}$ & $4.2\times10^{0}$    \\
\hline
3(4)$\sigma$1024S & sub & 1024 & 2    & $1.2\times10^{ 9}$ & $7.1\times10^{7}$
                  & $5.6\times10^{ 2}$ & $3.3\times10^{1}$    \\
3(4)$\sigma$512S  &  sub &  512 & 1    & $1.5\times10^{ 8}$ & $8.8\times10^{6}$
                  & $6.9\times10^{ 1}$ & $4.2\times10^{0}$   \\
3(4)$\sigma$128S  &  sub &  128 & 0.25 & $2.3\times10^{ 6}$ & $1.4\times10^{5}$
                  & $1.1\times10^{ 0}$ & $6.7\times10^{-2}$   \\
\hline
\hline
\end{tabular}
}
\caption{\small Spatial and mass resolution of the different top and
  sub grid simulations.  $L$ is the comoving box size of the top grid
  defining the longest wavelength perturbation, $\Delta x$ the
  comoving cell size in each computational box, $M_{DM}$ the total
  dark matter mass initially in each grid, $M_B$ the total baryonic mass,
  $\Delta M_{DM}$ the single dark matter particle mass, and $\Delta M_B$
  the baryonic mass in a single cell at the start of the simulation at
  redshift $z=100$.  The same parameters were used for both the 3 and
  4$\sigma$ perturbation simulations.
  We employ the notation 4$\sigma$1024T, for example, to denote the
  top grid simulation of a $4\sigma$  peak in the 1024$Mpc$  box.
 }
\label{tab:resolution}
\end{table}
For comparison, we note that the mass resolution is much smaller
than the Jeans mass in the background baryonic fluid,
which varies from about $10^6 M_\odot$ at $z=100$ to
$\sim 10^4 M_\odot$ at $z=10$.

\begin{figure}[htb]
\centerline{\psfig{file=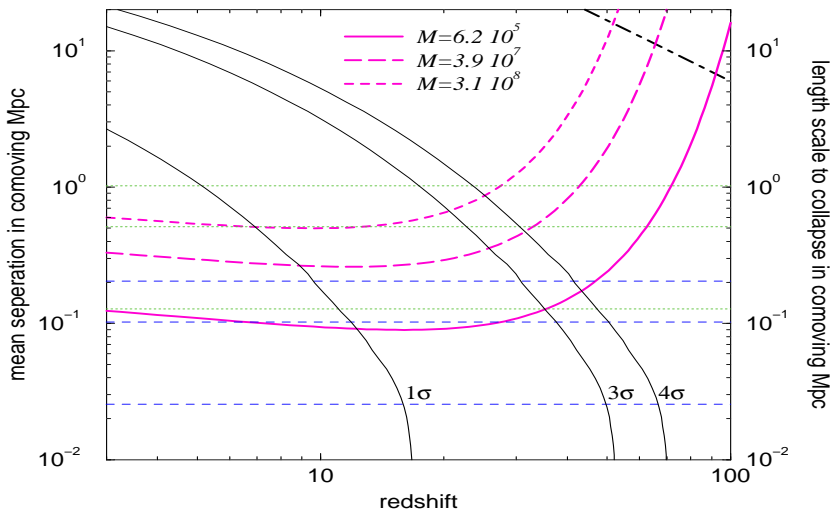,height=8cm,width=12cm}}
\caption{\small The solid lines labeled 1, 3, and $4\sigma$ depict the
 length scales of 1, 3, and $4\sigma$ peaks vs.  their collapse
 redshift as predicted by linear theory.  This assumes a Gaussian
 adiabatic CDM spectrum with $\sigma_8=0.7$, $\Omega_0=1$, and
 $h=0.5$, and a linear density contrast of $\delta_c=1.68$ at
 collapse.  The thick solid, long dashed, and dashed lines depict the
 comoving mean separation of peaks with a mass of $6.2\tento{5}$,
 $3.9\tento{7}$, and $3.1 \tento{8}\Ms$, as predicted by the
 Press--Schechter formalism, respectively.  These masses correspond to
 the three different perturbation wavelengths we have simulated:
 $25.6$, $102.4$, and $204.8\kpc$ (dashed horizontal lines).  The
 dotted horizontal lines show our box sizes, $1024$, $512$, and $128$
 comoving kpc.  For reference, the Hubble horizon is shown as a
 dot--dashed line (in the upper right corner).
}
\label{Fig:Lscales}
\end{figure}

In Figure~\ref{Fig:Lscales} we show the mean
comoving separation, computed from the Press Schechter (1974)
formalism, of the three different mass scales we have simulated.  The
solid, long dashed, and dashed lines correspond to a total collapsing
mass of $6.2\tento{5}$, $3.9\tento{7}$, and $3.1\tento{8}\Ms$
respectively.  These masses correspond to perturbation wavelengths
(dashed horizontal lines in Figure~\ref{Fig:Lscales}) of $25.6$,
$102.4$, and $204.8\kpc$, respectively.  On the same graph we also
plot the perturbation wavelength vs.  the redshift at which linear
theory predicts its overdensity to be 1.68 (its collapse redshift) for
$\sigma=1$, 3, and 4 peaks (solid downward sloping lines labeled
$1\sigma$, $3\sigma$, and $4\sigma$).  Our simulation box sizes
are indicated by the horizontal dotted lines.
We see that the $25.6\kpc$ perturbation
of the $4\sigma 128$ ($3\sigma 128$) simulation is expected to
collapse at a redshift $\sim 65$ ($\sim 50$).  At this redshift the
typical mean separation of such $6.2\tento{5}\Ms$ objects is $\sim
700\kpc$ ($\sim 220\kpc$) which is $5.5$ ($1.7$) times greater than
the simulation box size of $128 \kpc$.
For the larger box sizes (512
and $1024 \kpc$), the mean separation of peaks is larger than the box
at the collapse redshift by a factor of a few and about half the box
size for smaller redshifts.  Figure~\ref{Fig:Lscales} also shows that
at a redshift of $\sim 11$, $7$, and $5$ a ($1\sigma$) perturbation of
128, 512, and $1024\kpc$, corresponding to our box sizes, would go
non--linear, respectively.  Therefore, one cannot trust the simulation
results for smaller redshifts.  Although the mean separation of the
simulated peaks is always within a factor of five of the box sizes, the
use of periodic boundary conditions on these small scales will also
limit the reliability of our results.

\newpage
\section{RESULTS}
\label{sec:results}

\ssubsection{Morphology}

In Figure~\ref{cont_dens_top} we show an evolution sequence of the
logarithm of the dark matter and gas overdensities at four redshifts
$z=$ 35, 22, 17, and 12.  Five contour levels of $\log
\rho/\overline\rho = (0.5, 1, 1.5, 2, 2.5)$ are displayed in the
$x$--$y$ plane of the cube intersecting the maximum density peak.
Each plot in the figure represents the data from the top grid
evolution for the 3$\sigma$ collapse with the intermediate box size
$L=512\kpc$ ($3\sigma512$T). In the evolution sequence of the gas
density one can see two originally distinct objects (at $z=35$) that
merge later on.
By redshift $z=22$, a well developed spherical structure of
total mass $4\times10^7~M_\odot$ forms at the intersection of several
filaments, consistent with the prediction from linear theory that a
3$\sigma$ peak of mass $\sim 10^7~M_\odot$ will collapse at $z\sim
35$.  The structure continues to grow by the further collapse of
nearby particles and gas, and by the merging of smaller collapsed
structures.  The total mass of the bound structure grows to
$1.3\times10^8~M_\odot$, with baryonic mass $6.8\times10^6~M_\odot$ by
$z=12$.  Figure~\ref{cont_dens_sub} shows results analogous to those
in Figure~\ref{cont_dens_top}, but for the sub grid evolution.  The
enhanced resolution (a factor of four in length and 64 in mass) allows
for more detailed and accurate studies of the collapsing structure.
In fact, it is the enhanced resolution that allows the cooling flow to
be resolved adequately.  The top grid calculation does not have the
resolution required to produce significant amounts of hydrogen
molecules to cool the gas effectively.  Also, in the sub grid data, a more
complex network of filamentary structure is resolved due to the higher
mass resolution. Evidently the filamentary structures observed in the
gas appear also in the DM overdensity, consistent with the
picture that the baryons fall into the potential well of the dark
matter. In the gas overdensity (Figure~\ref{cont_dens_sub}) one
observes how a distinct overdense region forms in the upper filament
between redshifts 22 and 17, and merges then with
the more massive central structure by $z=12$. It is evident that a
significant amount of mass accretes onto the central object along the
filaments.

% Figure
% \centerline{\psfig{file=Figures/Subdens_dm_gas.ps,height=17cm,width=12cm}}
% here

Analogous to Figure~\ref{cont_dens_sub}, Figure~\ref{cont_temh2_sub}
shows the evolution sequence for the gas temperature and molecular
hydrogen number density fraction $n_{H2}/n_H$ on the sub grid.  The
contour levels are $\log T(eV) = -2.2, -2, -1.5, -1, -0.5$ and $\log
n_{H2}/n_H = -5.05, -5, -4.5, -4, -3.5$.
% Figure :
% \centerline{\psfig{file=Figures/Subtemp_h2.ps,height=17cm,width=12cm}}
% here
\HH fractions exceeding the background value of $\sim 3\tento{-6}$ are
found only in overdense regions along the filamentary structures, and
predominantly at their intersections where the densities are highest.
We find typical \HH fractions of $\sim 10^{-5}$
along the filaments, and
$\gsim 10^{-4}$ in the spherical overdense regions.  Both
of the spherical structures that are most evident
in the gas contour plots of Figure~\ref{cont_dens_sub}, and
merge between redshift 17 and 12, show \HH fractions exceeding
$3\tento{-4}$.  By redshift $z=12$ the fraction of H$_2$ in the
central structure increases to $9.1\tento{-4}$, which is enough to
cool the gas to $0.21eV~(2390K)$ and to collapse the gas to a central
density of $\sim100\cm^{-3}$.  Approximately 6\% of the baryonic mass
in the structure has cooled by $z=12$, so the collapsed and cooled
structure is roughly of mass $4\tento{5}\Ms$. We note that the merger
event did not destroy any molecules in the merging structures (see
also Figure~\ref{fig:fH2} below).

\ssubsection{Profile Plots}

In  the following  we  present radial  profile  plots which  represent
physical quantities   averaged  in spherical  shells  centered  on the
densest  cell in  the high resolution  sub grid.  All of these radially
averaged  quantities do not  include the central  zone.

\begin{figure}[htb]
\centerline{\psfig{file=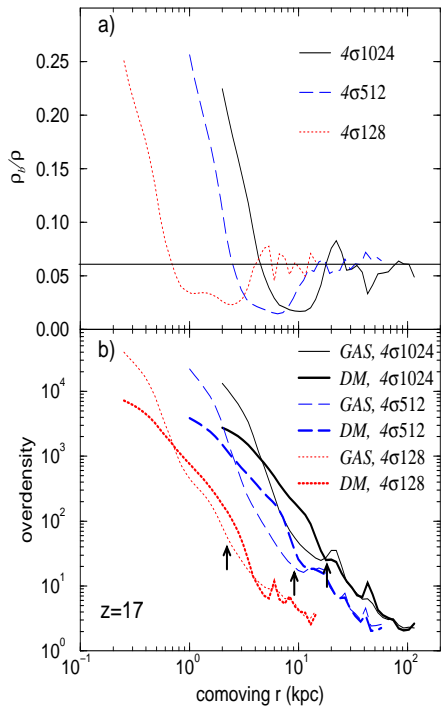,height=10cm,width=10cm}}
\caption{ (a) Radial profile of the baryonic fraction for the
  $4\sigma$ runs.  The horizontal solid line is the average baryonic
  fraction $\Omega_b=0.06$.  (b) Radial profile of gas and dark matter
  overdensities for the $4\sigma$ runs.  Note that the baryonic
  density increases substantially relative to that of the dark matter
  in the central core due to $H_2$ cooling. The arrows indicate the
  theoretical virial radii of the three different perturbation
  wavelengths we have simulated.  }
\label{Dvsr}
\end{figure}

Figure~\ref{Dvsr}a shows the  ratio  of baryonic to total  density for
all of the $4\sigma$ runs vs.  the distance from  the center in comoving kpc.
The horizontal line  depicts    the  background (initial)   value   of
$\Omega_B/\Omega = 0.06$.  Evidently the cooling  at the center of the
structure allows the baryons to contract, leading to an enhancement of
baryons  over the dark matter up  to four  times the background value.
The radius  at which the ratio of  baryonic to total density starts to
exceed $\Omega_B/\Omega = 0.06$ coincides with the radius at which the
molecular hydrogen fraction  reaches $5\tento{-4}$.  This value agrees
well with the analytical  model presented in  section~\ref{sec:AModel}
Note that the virial shock is located at radii larger than where the
deficit is found. Hence, it is the shocked gas that forms molecules,
cool, and settles to the center of the structure yielding the 
baryonic to DM bias.
(see also Figure~\ref{fig:fH2}).  The cooling gas flows from the outer
region to the center  causing the the  deficit of baryons compared  to
the background value seen in Figure~\ref{Dvsr}a.

The baryon enhancement at the center of the structures leads to sharp
overdensity profiles as shown in Figure~\ref{Dvsr}b.  The radially
averaged dark matter density profile shows a complex form with
apparently four regimes of different slopes.  Powerlaws that fit this
radial dependence have an exponent $\lsim -1$ at the central regions,
then $\sim -2$, $-4$, and again $-2$ at the outer regions.  The arrows
in Figure~\ref{fig:fH2} indicate the theoretical viral radii of
collapsed spheres with intial wavelengths corresponding to the ones we
have simulated ($r_{vir} = (l/2)(18\pi^2)^{-1/3}$, where $l$ is the
pertubration wavelength).  Note that for the least massive halo the
$r^{-4}$ feature is found just outside the virial radius, where for
the others its partially ($4\sigma512$) and completely ($4\sigma1024$)
within the virial radius. This is because we plot the profiles at
redshift 17 at which only the smallest simulated structures is fully
virialized. Generally we see the halo DM profiles within the virial
radius to be isothermal ($\rho \propto r^{-2}$) over a wide range of
radii.  This shape is the typical behavior of the universal density
profile proposed by Navarro, Frenk, and White (1996, NFW hereafter),
\begin{equation}\label{equ:NFW}
  \rho(r) \propto \frac{1}{(r/r_s)(1+r/r_s)^2,}
\end{equation}
where $r_s$ is some characteristic radius.  They showed this formula
to fit DM haloes within the virial radius of the galaxies and clusters
of galaxies found in their N--body simulations.  Although Navarro
\etal studied DM profiles for halos with masses $\sim 10^{12} \Ms$,
their results are similar to what we find in our simulated halos,
which are approximately $10^6$ times smaller in mass.  However, our DM
profiles are somewhat sharper than the NFW profile making a future
high resolution study desirable to test the applicability of the
universal profile to DM halos on mass scales $\lsim 10^8\Ms$ and
clarify the ``feedback'' of the collapsing baryons on the DM density
profile.

\begin{figure}[htb]
\centerline{\psfig{file=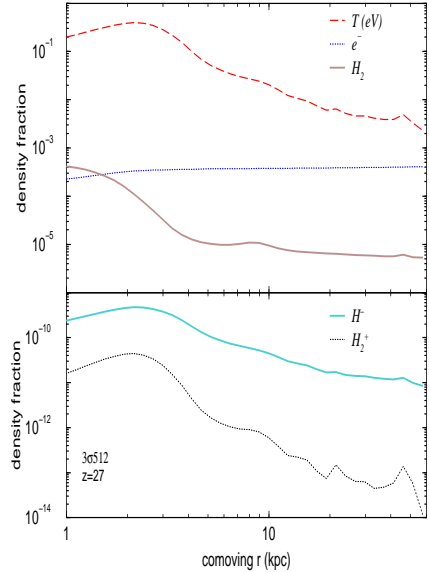,height=8cm,width=6cm}}
\caption{  Profiles   of temperature   and  several important  species
  (H$_2$,  H$_2^+$, H$^-$ and $e^-$)  at  $z=27$ for the  $3\sigma512S$
  run.  The depletion of \Hm,  \HHp, and e$^-$  in the \HH forming and
  cooling region is evident.  The virialization shock is not
  ionizing (yet) since the free electron  fraction remains constant away
  from the central regions.}
\label{all.3_512.z27}
\end{figure}

\begin{figure}[htb]
\centerline{\psfig{file=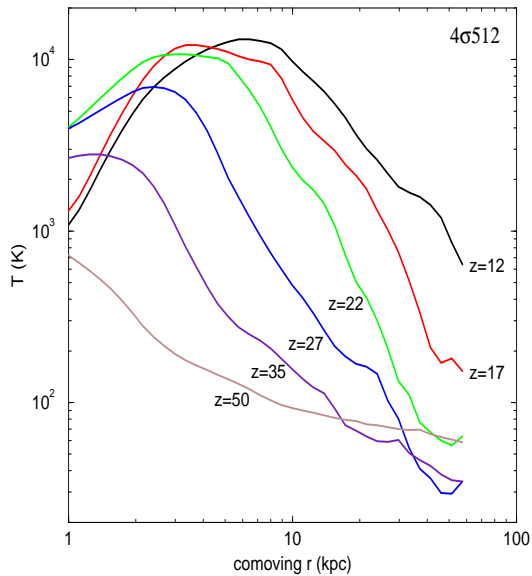,height=8cm,width=8cm}}
\caption{ Radially averaged temperature profiles for the
  $4\sigma512$S sub grid evolution.  The adiabatic compression of the
  central regions is the dominant effect at redshift $z=50$.  At
  $z\sim35$, H$_2$ cooling dominates
  adiabatic heating in the central--most regions,
   and effectively cools the gas. At redshifts
  $\lsim 25$ the radially averaged temperature exceeds $10^4\K$ at a
  few comoving kpc from the center.}
\label{Tvsr}
\end{figure}

Figure~\ref{all.3_512.z27}  illustrates that  the  central regions of
collapsing  small  scale structure  can cool  with  molecules that are
formed using free electrons that  were left over from recombination as
catalysts. It shows the radially  averaged profiles for the fractional
abundances of \HH, \Hm, \HHp,  and e$^-$, as  well as the  temperature
for the $3\sigma512$S data at  $z=27$.
The molecular hydrogen formation is dominated by the \Hm path, which is
similar to the situation of strong shocks ($v_s\sim 100\kms$) formed
during the collapse of cosmological sheets (Anninos \& Norman 1996,
and Abel \etal 1997).
This is clear from Figure~\ref{all.3_512.z27} since the
\Hm abundance exceeds the \HHp abundance by always more than an order
of magnitude, and the fact that the dissociative attachment reaction
of \Hm to form \HH is characterized by a rate coefficient that is two
times greater than the charge exchange reaction of \HHp with H
(reactions (\ref{reac:H2PH}) and (\ref{reac:HMH})).  The free electron
abundance is decreased in the more dense and cooling center where the
radiative recombination timescales are short.  The weak shock
is not capable of ionizing the primordial gas at this
redshift.  The decrease of the \Hm abundance in the cooling layer is a
result of the temperature decrease and the free electron depletion.
Note that these results justify all the assumptions that entered the
derivation of \eq{equ:fH2}, further validating the analytical
description given in section~\ref{sec:AModel}.  Although the central
parts of the structure formed hydrogen molecules solely from the
residual free electrons as catalysts, at later redshifts the infalling
material will be, at least partially, collisionally ionized by the
stronger shock, allowing molecules to form on a faster timescale (see
Figure~\ref{Tvsr}). Furthermore, we can see that the minimal
chemistry model proposed by Abel \etal (1997a) would be sufficient to
study the effects of primordial gas chemistry for first structure
formation.

\ssubsection{Evolution}

The temperature profiles for different redshifts of the $4\sigma512$S
box are shown in Figure~\ref{Tvsr}.  (The central temperatures can be
read from Figure~\ref{TestADIA}.)  Clearly the virialization shock
becomes stronger as more baryons collapse towards lower redshifts.  It
is interesting that for the small scale structures considered here
that the central region never reaches temperatures above a few
thousand Kelvin. For the $4\sigma128$S run the maximum temperature in
the center of the densest structure is only $\sim 900\K$ (see
Figure~\ref{TestADIA}).  At $z= 50$, the temperature profile shows
a gentle increase towards the center caused by adiabatic compression
with its maximum at the center. At $z= 35$, the influence of
radiative cooling causes the maximum temperature in the structure,
which increases steadily through mass accretion, to
be found away from the high density center core. At redshifts $z\lsim22$, the
radially averaged post--shock temperature exceeds $10^4\K$, and hence
the gas becomes collisionally ionized, and more and more
molecules form in the center to cool the gas to lower and
lower temperatures.

\begin{figure}[htb]
\centerline{\psfig{file=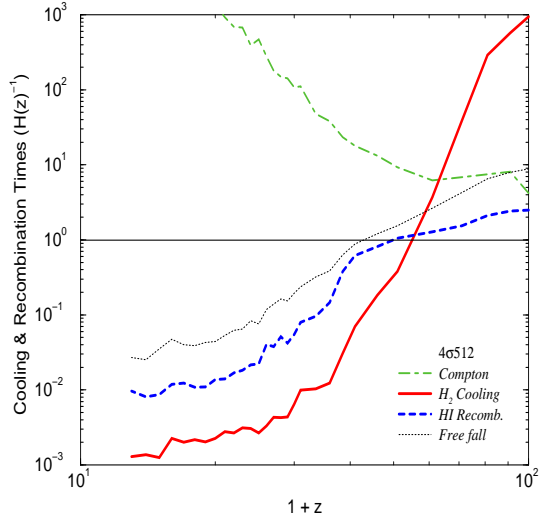,height=8cm,width=8cm}}
\caption{Cooling times in the densest cell found on the sub grid of
  the simulation $4\sigma512$S, relative to the Hubble time
  $H^{-1}=H_0^{-1} (1+z)^{-3/2}$, for various physical processes, including
  Compton, and molecular hydrogen line cooling.  Also shown are the
  hydrogen recombination and spherical free fall times.}
\label{fig:evol_cool}
\end{figure}
Figure~\ref{fig:evol_cool} shows some
characteristic time scales normalized to the
Hubble time ($H^{-1}=H_0^{-1} (1+z)^{-3/2}$, where $H_0$ is the
present Hubble constant) of the most important physical processes in
the densest cell of the collapsing structure in the $4\sigma512$S
simulation.  The Bremsstrahlung and recombination cooling times exceed
the expansion time scale by more than three orders of magnitude and
are therefore not shown in the Figure. The hydrogen
recombination time is already close to the Hubble time
at $z=100$, allowing \HH to form
slowly. At redshift $z\approx 50$, the gas density and temperature
grow to the levels that are more favorable for the creation of
hydrogen molecules and the H$_2$ cooling time becomes shorter than the
hydrogen recombination (\eq{equ:trec}) and dynamical ($t_{dyn} =
(3\pi/(32 G\rho))^{1/2} = 4.7\tento{7}n^{-1/2} \yrs$) times.
The Compton cooling/heating exceeds the expansion
time scale for all redshifts~$\lsim100$ and does not influence the
collapse of the central regions. The reason that the \HH cooling time
is monotonically decreasing with redshift and does not start to
increase again is due to the merging of small scale structures and, for the
largest box sizes, an artifact of our limited numerical resolution as
we will show in the following.

\begin{figure}[bht]
\centerline{\psfig{file=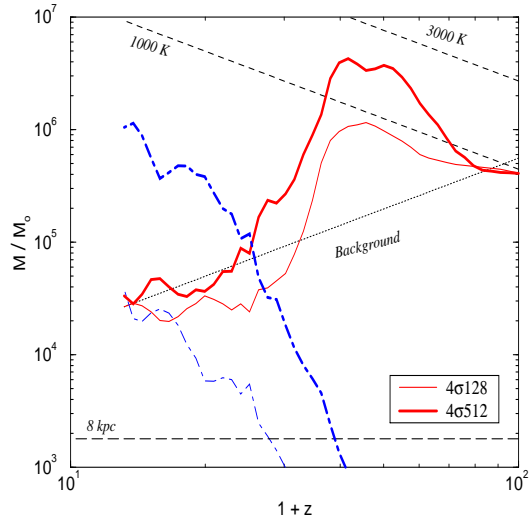,height=8cm,width=8cm}}
\caption{Evolution of the Jeans mass in the densest cell of the most
  massive structures in the $4\sigma128$S and $4\sigma512$S
  simulations (thin and thick solid lines).
  Also shown are the corresponding mass in
  the densest cell (thin and thick dot--dashed lines), the Jeans mass of the
  background baryonic density (dotted line), the initial baryonic mass
  in a single cell of the coarsest grid simulation (the parent 1024
  kpc grid with a comoving cell size of 8 kpc, horizontal dashed
  line), and the Jeans mass for spherical collapse with virial
  temperatures of 1000 and 3000K (dashed lines labeled with 1000K and
  3000K, respectively).}\label{fig:evol_jmass}
\end{figure}

\begin{figure}[hbt]
\centerline{\psfig{file=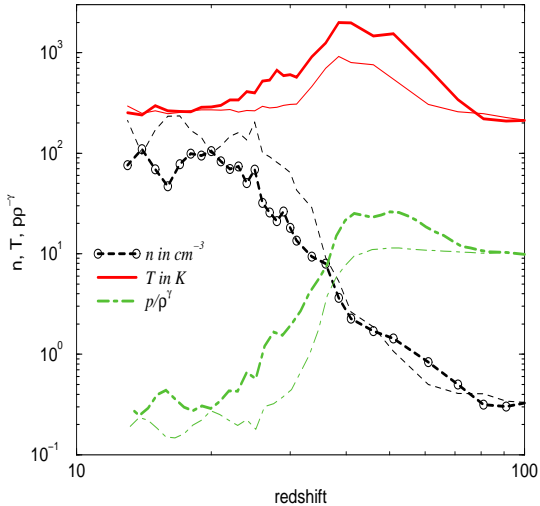,height=8cm,width=8cm}}
\caption{The evolution of  the baryonic  number density in  $\cm^{-3}$
     (dashed  lines    with circles  indicating    the  data  points),
     temperature in degrees Kelvin (solid lines), and
     the ``entropy'' function $p/\rho^\gamma$ which is
     invariant    for  adiabatic   processes  (dot--dashed   lines) in
     arbitrary units for the densest zone in the simulation box of the
     $4\sigma128$S (thin lines) and  $4\sigma512$S (thick lines) data.
     Here $p$, $\rho$, and $\gamma$ denote  the gas pressure, density,
     and the adiabatic index, respectively.}
\label{TestADIA}
\end{figure}

The decrease in temperature and increase in density at the cores
results in an even greater decrease in the Jeans mass
\begin{equation}
M_J = \frac{\pi}{6} \rho \lambda_J^3 =  \frac{\pi}{6} \rho \left(
\frac{\gamma \pi k T}{G \rho \mu m_H} \right)^\frac{3}{2} \approx
 100 T_K^{3/2} n_B^{-1/2} ~M_\odot ,
\end{equation}
as shown in Figure~\ref{fig:evol_jmass}.
Here $n_B$ denotes the
baryonic number density expresses in numbers of hydrogen atoms.  The
background Jean's mass, shown in
Figure~\ref{fig:evol_jmass}, is derived from the background baryon
number density $1.1 \tento{-5} (\Omega_B/\Omega_0)(\Omega_0 h^2)
(1+z)^3 \cm^3$).  Also shown in Figure~\ref{fig:evol_jmass} is the
initial mass resolution in the coarsest grid (the parent grid with
$L=1024$ Mpc) simulations with a cell size of 8 kpc, the baryonic mass
in the single densest cell of the collapsing structures, and the
Jean's mass computed using the spherical collapse model with virial
temperatures of $1000\K$ and $3000 \K$,
\begin{equation}
M_S = 100 T_K^{3/2}
  \left(18\pi^2 n_0 \right)^{-1/2} (1+z_c)^{-3/2} ~M_\odot ,
\end{equation}
where $z_c$ is the collapse (virialization) redshift. The redshift at
which the single cell mass curves cross the corresponding Jeans mass
curves represents a limit below which the grid resolution becomes
inadequate to resolve the cooling flows. In the
larger mass structures ($512$ and $1024\kpc$ box sizes), the
collapsed mass exceeds the Jeans mass of the cell at higher
redshifts.  The horizontal dashed line showing the initial mass
resolution of the coarsest grid indicates that the mass resolution in
the simulations is more than sufficient to track the baryonic Jeans
mass up to the single cell mass crossing time.
The minimum Jeans mass reached in the present simulations
($\sim 3\tento{4}~M_\odot$) suggests that the insufficient spatial resolution
does not allow any predictions about fragmentation and the primordial
star formation process directly.  However, the physics of the initial
evolution and the physical environment of primordial star forming
regions are adequately captured in our numerical experiments. Clearly,
higher resolution studies are desirable.

\ssubsection{Comparison to the Analytic Model}
\label{sec:comparison}

In the following we will test some of the analytical arguments
introduced in section~\ref{sec:AModel}.  One typical assumption that
is made (see Tegmark~\etal 1997), is that a structure evolves
adiabatically until virialization.  However, at small wavelengths, the
CDM perturbation spectrum is almost flat causing these structures to go
nonlinear almost simultaneously in time and one expects their evolution
to be marked by frequent mergers which could raise the gas to a higher
adiabat by shock heating.
This is what is illustrated in Figure~\ref{TestADIA} which shows the
evolution of the ``entropy'' $p/\rho^\gamma$ (which is constant for adiabatic
processes), the temperature, and the baryonic number density of the
densest zone of the $4\sigma128$S (thin lines)
and $4\sigma512$S (thick lines) data.  Clearly for
the $4\sigma128$ simulation, the initial collapse at
$z\gsim 45$ is adiabatic ($p/\rho^\gamma=const.$) and radiative
cooling plays a role only at later redshifts.  For the larger
$4\sigma512$ simulation box, the entropy increases in the
redshift range $60 \lsim z \lsim 75$ due to merging and stays roughly
constant thereafter
until radiative cooling becomes important at $z \lsim 40$.
\HH cooling gives rise
to a plateau in temperature until $z\lsim 38$ when it
causes the temperature to decrease.
Interestingly, we see that the central regions never reach the
theoretical virial temperatures which are $\sim2600\K$ for the
$4\sigma128$, and greater than $10^4\K$ for the $4\sigma512$ run.
Our simulations fail to resolve the core of these structures
and hence the central density is not well defined once the gas in the
most central structure has cooled.  Therefore, the slow growth of the
density for redshifts $10<z<30$ does {\em not} indicate that the
collapse is halted.

The post--shock gas has been found to evolve isobarically in the case
of radiative shocks (see, for example, Shapiro \& Kang 1987 and
Anninos \& Norman 1996).
Assuming the molecular hydrogen fraction increases
according to \eq{equ:fH2} (see also Figure~\ref{fig:fH2}) and the
isobaric evolution equation as discussed in section~\ref{sec:AModel}, one
can determine the temperature evolution.  For the central
(densest) zone of our
$4\sigma128$S box, this isobaric evolution predicts that the gas would
be able to cool to $\sim 30\K$ by redshift ten.  The Lepp and Shull
cooling function, which was used to derive the isobaric evolution, has
been stated to be only valid to $100\K$.  However, a new study by
Tin\'e, Lepp, and Dalgarno (1997) confirms that the \HH cooling
timescale can be short even at lower temperatures.  For example they
find it to be of order $1\Gyr$ at a temperature of $50\K$ and a
tenth of that at $70\K$ for typical \HH fractions of $10^{-3}$ and
$n_H \sim 100 \cm^{-3}$.  Hence if the gas evolves isobarically
it will be able to cool to temperatures $\ll 100\K$. However, we see
that the temperature and density, and hence the internal energy stay
roughly constant at the final stages  although the cooling time scales
are short. 
This is due the extensive
merging in the hierarchical scenario considered here, as well as our
limited numerical resolution.

In section~\ref{sec:AModel} we discussed the simple estimate of Abel
(1995) and Tegmark \etal (1997) for the evolution of the \HH fraction
in a cloud derived solely in terms of its virial temperature and initial
recombination timescale.
\begin{figure}
\centerline{\psfig{file=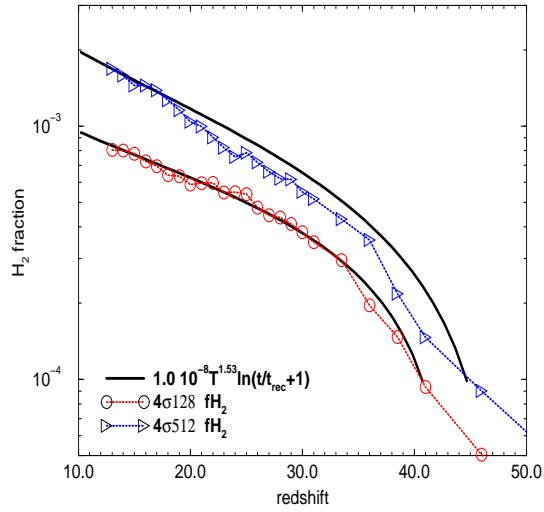,height=8cm,width=8cm}}
\caption{Molecular hydrogen fraction vs. redshift in the densest
  zone of the $4\sigma128$S (open circles) and $4\sigma512$S
  (triangles) simulations. The thick solid lines depict the approximating
  \eq{equ:fH2} initialized with the appropriate free electron
  densities, initial \HH fraction, and temperature.}
\label{fig:fH2}
\end{figure}
In Figure~\ref{fig:fH2} we  compare the predictions
of \eq{equ:fH2} with the results
from    the  $4\sigma128$S  (open    circles)  and  the  $4\sigma512$S
(triangles) runs.  The fit is remarkably good, although the initial
temperature  (or redshift) is  somewhat difficult to  pick out in this
case since  the heating due to  adiabatic compression is  slow and the
initial (virial) temperature is not as well defined  for
the larger mass  scales.  Analyzing the time derivative of
\eq{equ:fH2} one finds  for  large times that   $\dot{f_{H_2}} \propto
T^{1.53}$ which,  for  the  spherical collapse  model,   translates to
$\dot{f_{H_2}} \propto  M (1+z_{vir})^{1.53}$.  This explains that the
divergence of the two graphs in Figure~\ref{fig:fH2} for low redshifts
is due to the difference in the collapsing mass and collapse redshift.

Tegmark \etal (1997) have recently used a simple approach to derive
the  minimum mass  scale that is  able  to collapse  as a function  of
redshift.  They model the density evolution  in a cloud according to
the  predictions of the  spherical   collapse model, assuming the
temperature increases adiabatically up to
virialization and changes thereafter by radiative cooling.
They simultaneously  solve the rate
equations  for the time  dependent chemistry to accurately predict \HH
formation  and the subsequent cooling of the gas.
We have reproduced their work with
the   same cooling function   (Lepp  \& Shull 1983, hereafter referred
to as LS)    used in  our
cosmological hydrodynamics code to  be able to directly compare  their
findings to our 3D numerical results.
\begin{figure}[htb]
\centerline{\psfig{file=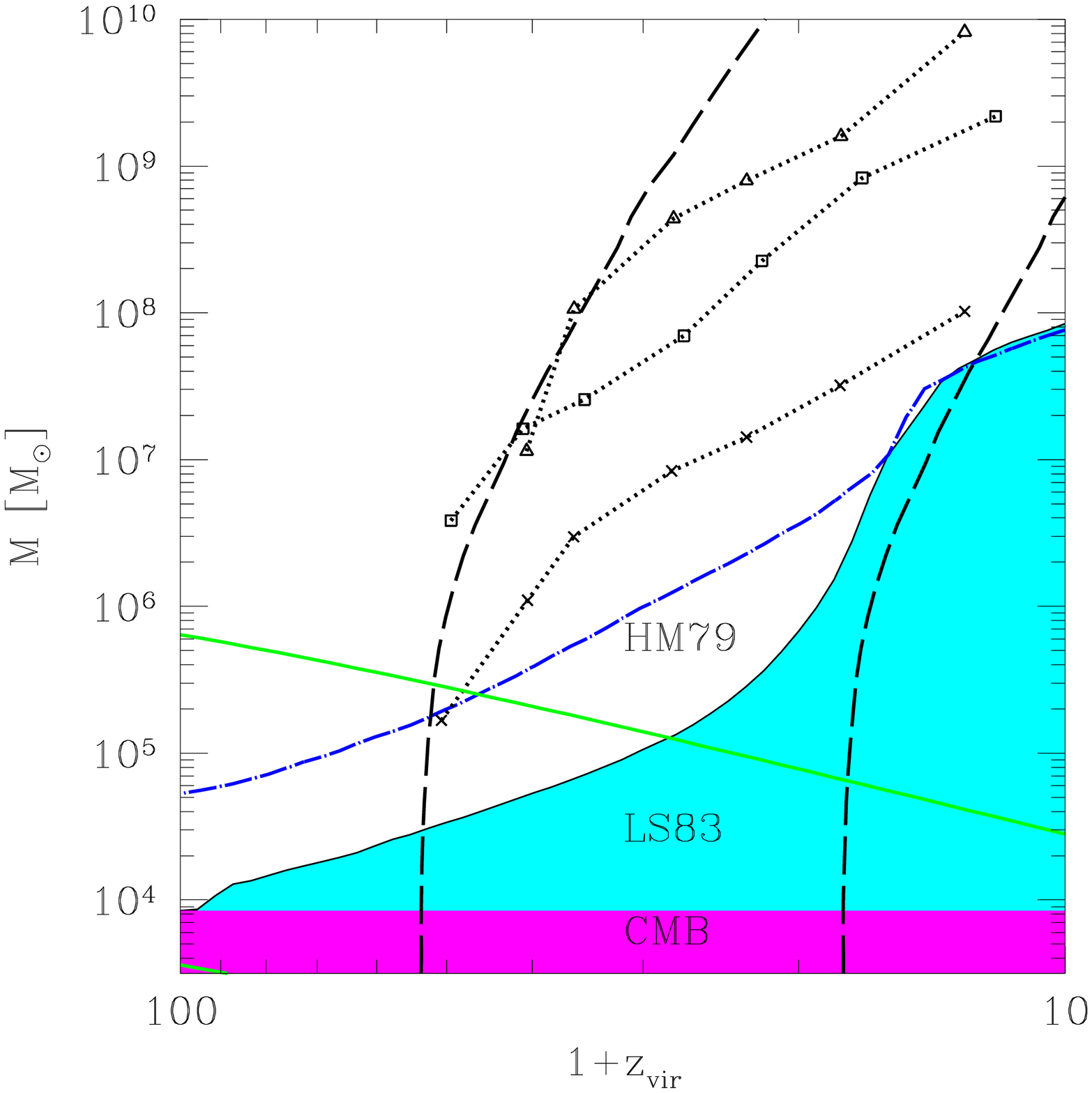,height=8cm,width=8cm}}
\caption{
Total (baryonic and dark matter)  collapsed
mass vs. redshift.  The horizontal
dark shaded region depicts the mass scale for
         which the virial temperature equals the CMB temperature.
The light shaded area (labeled LS83) represents the domain of parameters
for which structures cannot collapse. This curve is computed
assuming a spherical collapse model as in Tegmark \etal (1997), except
we use the Lepp \& Shull (1983) \HH cooling function.
         Only above the light shaded area are
         structures believed to be able to collapse.
         The dotted lines show $M_{200}$ from our $3\sigma128$S (crosses),
         $3\sigma512$S (squares) and $3\sigma1024$S (triangles) runs. The two
         dashed upward sloping lines depict the collapse redshift
         given by linear theory for the standard CDM spectrum scaled
         appropriately for $1\sigma$ and $3\sigma$ perturbations. The
         gray solid downward sloping line depicts the Jeans mass at
         $18\pi^2$ times the background density.  The dot--dashed line
         is the original delimiting line computed by Tegmark
         \etal based on a modified form of the
         Hollenbach and McKee (1979) \HH cooling function.  }
\label{Teg96}
\end{figure}
Figure~\ref{Teg96} is  analogous to Figure  6 of Tegmark  \etal (1997)
with  our numerical results  superimposed.  The dotted lines are found
by adding  up all the  mass in the $3\sigma$  simulations
that is found  within spherical shells on
the sub grid in which  the mean enclosed  overdensity  is 200.
The collapse redshift agrees reasonably well with
the predictions of linear theory, represented by the two
dashed upward sloping lines.
The use of  different cooling  functions has a very strong
influence  on the mass  that can  collapse, as evidenced by
the difference in the two regions labeled HM79 (the Tegmark \etal
result) and LS83 (our result).
However, we note that although the quantitative
results are very different, the shape for these two different $M_c(z)$
curves is  rather  similar at  redshifts  $30<z<100$.  Their  slope is
consistent with  $M_c  \propto  (1+z)^{-3/2}$,  indicating  a constant
virial temperature  since  $T_{vir} \propto  M^{2/3}  (1+z)$.  For the
case in  Tegmark \etal, the  virial  temperature in  that
redshift interval
($30<z<100$) is approximately $1000\K$. A constant
virial temperature, in turn,
implies a constant final \HH fraction given by \eq{equ:fH2} which
yields $f_{H_2} \sim 5 \times 10^{-4}$ for $T_{vir}\sim 1000\K$ 
Comparing the cooling time scale to the dynamical time, they further
argued that this \HH fraction represents a universal value for which,
if exceeded, the cloud will collapse at the free fall rate. This
argument depends strongly on which cooling function is assumed.  Using
the Lepp and Shull (1983) \HH cooling function, we find that the
 minimum virial
temperature needed to produce enough molecules so that the gas can
cool faster than its dynamical time is $\sim 200\K$ for redshifts
$>30$. Which is five times smaller than the value derived from the
modified Hollenbach and McKee (1979, hereafter HM79)
cooling function
used by Tegmark~\etal.  This smaller minimum virial temperature for
the Lepp and Shull cooling function
translates to a critical \HH
fraction of only $\sim 3 \times 10^{-5}$.  In our simulations, we find
that the radius at which the baryonic overdensity exceeds the DM
overdensity (see Figure~\ref{Dvsr}) is coincident with the radius at
which the fractional abundance of \HH molecules exceeds $\sim
5\tento{-4}$. That we recover the critical \HH fraction computed by
Tegmark \etal, although we used the Lepp and Shull cooling function in
our simulations, is hence coincidence.

\section{Discussion and Summary}
\label{sec:discussion}

We have performed  several three--dimensional numerical simulations of
the  first bound objects  to form in   a CDM model  from high $\sigma$
fluctuations.  In addition to  evolving  the dark and baryonic  matter
components,  we have also  solved  a reaction  network  of kinematical
equations for the chemistry   important to the production  of hydrogen
molecules.    The coupled system  of  chemical reactions and radiative
cooling have been solved  self--consistently, without recourse  to any
equilibrium or other simplifying assumptions. The accurate modeling of
non--equilibrium cooling is  crucial in  determining the abundance  of
hydrogen molecules that  form in the  cores of  collapsed objects.   A
range of  baryonic mass  scales $10^{3}$ --  $10^{8}\Ms$  (set  by the
range of box sizes) and formation epochs $15 \lsim z \lsim 70$ (set by
both  the  mass  scale  and   the normalization  $\sigma$)  have  been
investigated in this paper.   The baryonic  gas  evolves to a  complex
network  of   sheets,  filaments  and  knots    with typical  baryonic
overdensities in the filaments of roughly an  order of magnitude lower
than at their intersections where $\rho/\overline\rho > 200$.
At all  epochs and mass scales
covered in our  simulations, the   heated  gas in  the filaments  have
characteristic temperatures that are  typically a few times lower than
in the spherical  knots.  Although we find  the  abundance of hydrogen
molecules   is enhanced   throughout all the   overdense or  collapsed
structures, the lower    densities and temperatures of  the  filaments
result  in the  lower   fractional  abundances of  $n_{H2}/n_H   \lsim
5\tento{-5}$ along the  filamentary structures as compared  to typical
concentrations  of $\gsim 5\tento{-4}$  in  the spherical knots at the
intersections of the filaments.
This is understood from the analytical \HH fraction evolution given in
\eq{equ:fH2} which implies that hydrogen molecules form at slower
rates in gas with lower initial temperatures and baryon densities.
Since the \HH cooling timescale in the low--density limit is
proportional to $1/(n_H f_{H_2})$, it is clear why significant cooling
is observed only at the cores of the highest density spheroidal
structures, and not along the filaments.
On the other hand, most of the volume in the simulations
consists of expanding underdense voids. In these regions,
the \HH fraction
equals the primordial background value of $\sim 3\tento{-6}$ since the
densities and temperatures there are too low to further produce \HH.
Our findings are consistent with the results of Tegmark~\etal~(1997)
who used a simple spherical description of primordial collapsing
clouds to estimate the minimum mass scale that can collapse for a
given redshift.  We have also justified the findings of Abel (1995)
and Tegmark \etal (1997) that the molecular hydrogen fraction formed
in structures at high redshifts arises initially solely through the
residual free electrons left over from the incomplete recombination of
the universe.  Furthermore, our numerical results confirm the rule of
thumb derived by Tegmark \etal that can be stated as: ``Once the \HH
fraction is $\gsim 5\tento{-4}$ in a spherical primordial gas cloud
and the gas temperature is well above $100\K$, the gas will be able to
cool on timescales $\lsim$ its dynamical one.''

We have not observed hydrogen molecules to trigger the initial
collapse of perturbations at high redshifts.  The adiabatically
compressed core densities and temperatures in the initial phases of
collapse are not high enough for the chemistry to generate a large
abundance of hydrogen molecules.  The gas therefore cannot cool before
it collapses gravitationally and shock heats to temperatures greater
than a few hundred degrees Kelvin.  Our results thus confirm the
conclusions of Haiman \etal (1996b) who found, using one--dimensional
spherical models, that radiative cooling by \HH affects the collapse
of the baryonic gas only after it has fallen into the potential well
of the dark matter halos and virialized.  We note that Haiman \etal
(1996b) overestimate the background \HH fraction since they
underestimated the photo--dissociation of \HHp molecules at high
redshifts through the CBR.  This overestimation of the \HH fraction in
the background primordial gas, by two orders of magnitude, causes them
to find that structures with virial temperatures as small as $120\K$
would be able to cool via \HH.  Consequently, their results on the
history of \HH formation in the virialized system cannot be correct.
However, since they find that \HH molecules cannot trigger the
collapse of primordial structure despite the overestimated \HH
fraction, this result, which has been confirmed in this study, is
robust.  Although \HH cooling does not trigger the collapse, it
significantly influences the time evolution by keeping the
temperatures in the central regions of the spherical knots to below
the theoretical virial temperatures by at least a factor of a few, and
by enhancing the baryon overdensities to $>10^4$.  Furhermore we find
\HH cooling to cause a typical gas to DM density bias in the central
regions similar to 4 times its background value.

\begin{figure}
\centerline{\psfig{file=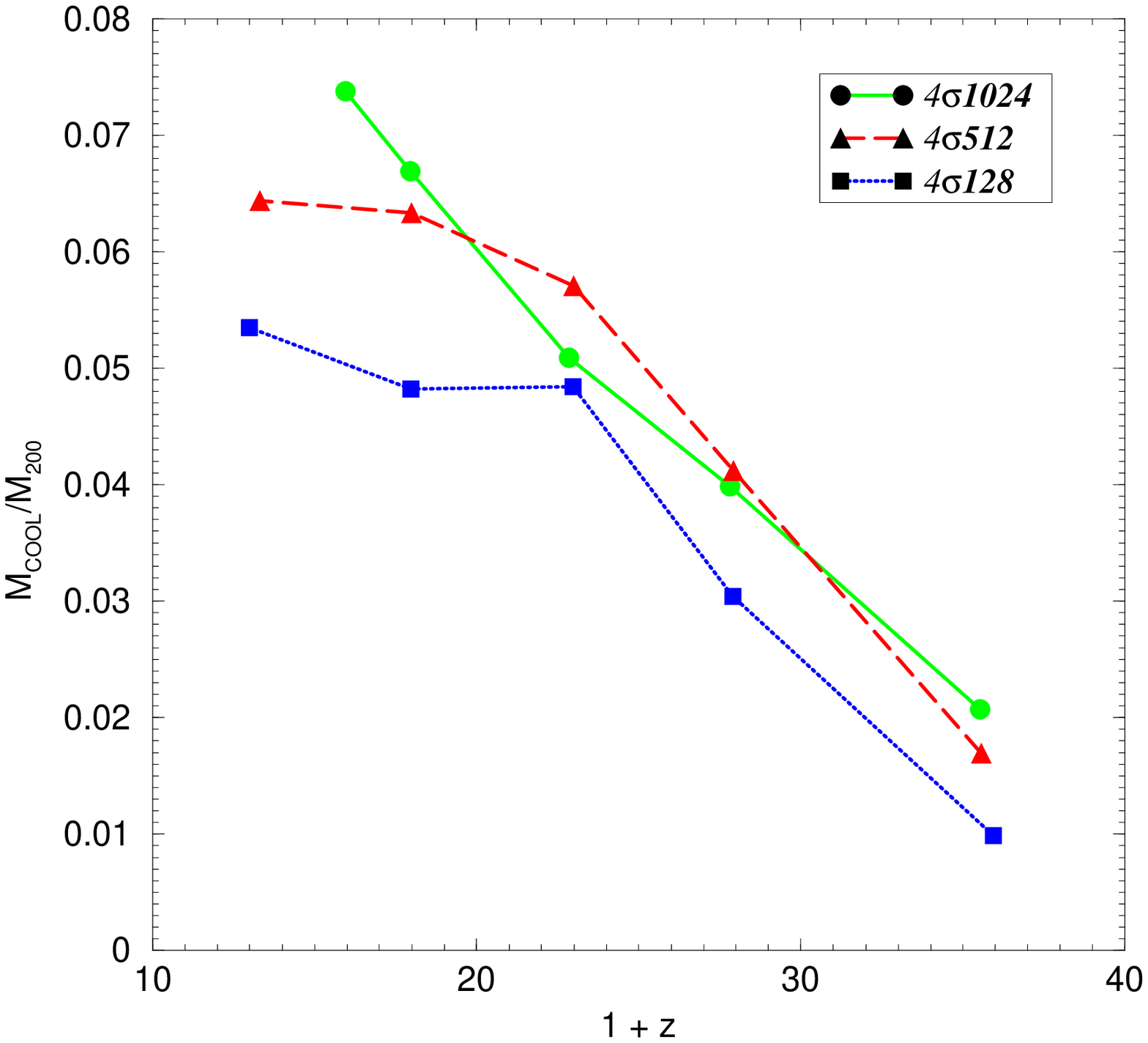,height=8cm,width=8cm}}
\caption{Evolution of the ratio of the cooled baryonic mass to the
  baryonic $M_{200}$ is  shown for the 3 different  (128, 512, \& 1024
  kpc box  size)  $4\sigma$ runs.   The fraction of gas which cools
  approaches 5 to 8\% towards the lower redshifts.
The apparent slower growth   of  the structure in   the  $4\sigma1024$S
simulation is an artifact of grid resolution since the cooling flow is
not resolved as well as in the smaller grids.
  }
\label{M_coolvsz}
\end{figure}

\begin{figure}[thb]
\centerline{\psfig{file=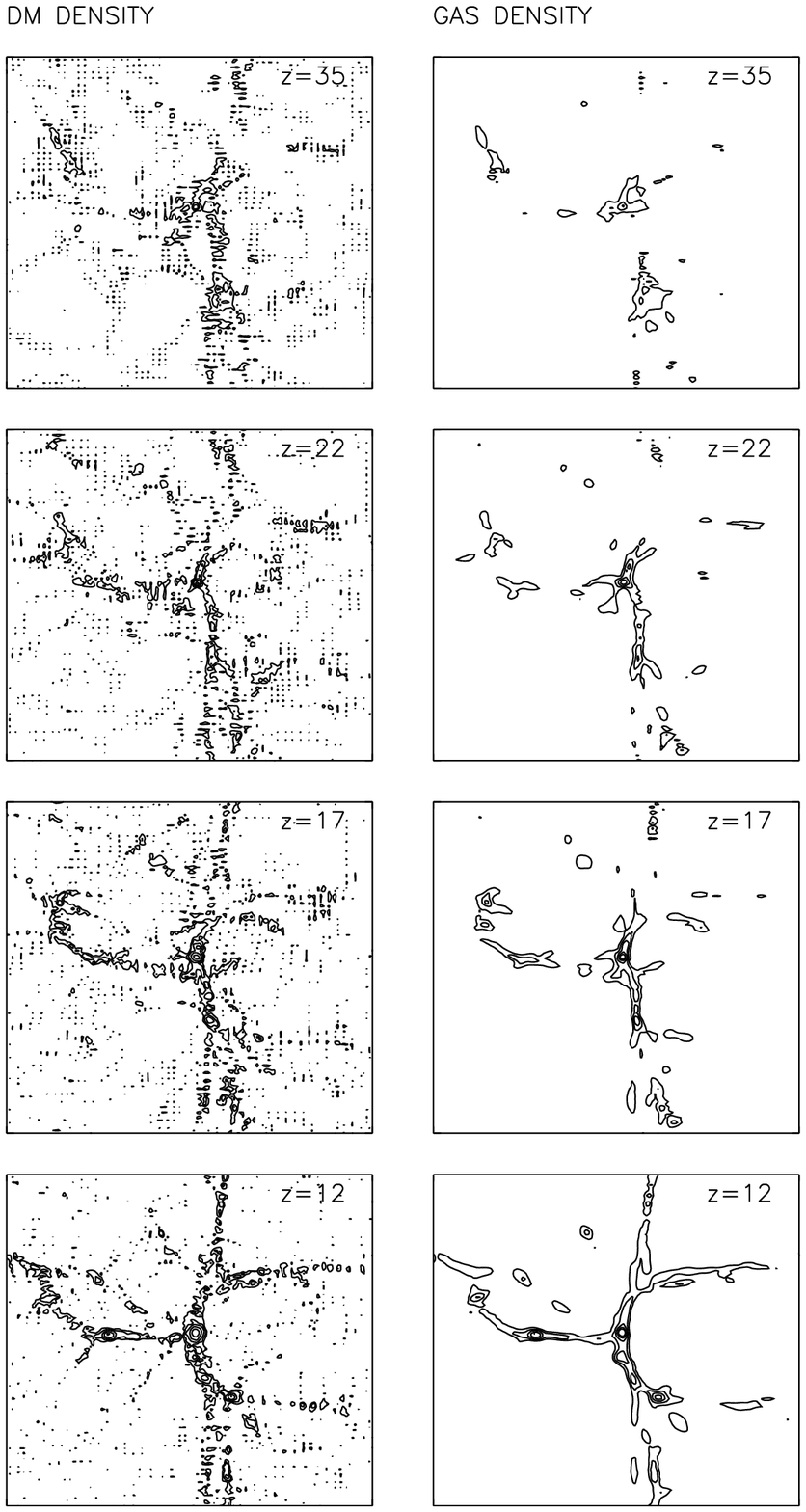,height=17cm,width=12cm}}
\caption{
  Contour surfaces  showing an evolution  sequence for the dark matter
  and baryonic gas overdensities at redshifts $z=$  35, 22, 17 and 12.
  Five   levels  (0.5,  1, 1.5,  2,    2.5) are  displayed   for $\log
  \rho/\overline\rho$,  where   $\overline\rho$  is    the  background
  density. Each individual plot represents a single slice at a fixed $z$
  coordinate, intersecting  the densest cell on  the top grid  for the
  3$\sigma$ run with box size $L=512~kpc$.  }
\label{cont_dens_top}
\end{figure}

\begin{figure}[thb]
\centerline{\psfig{file=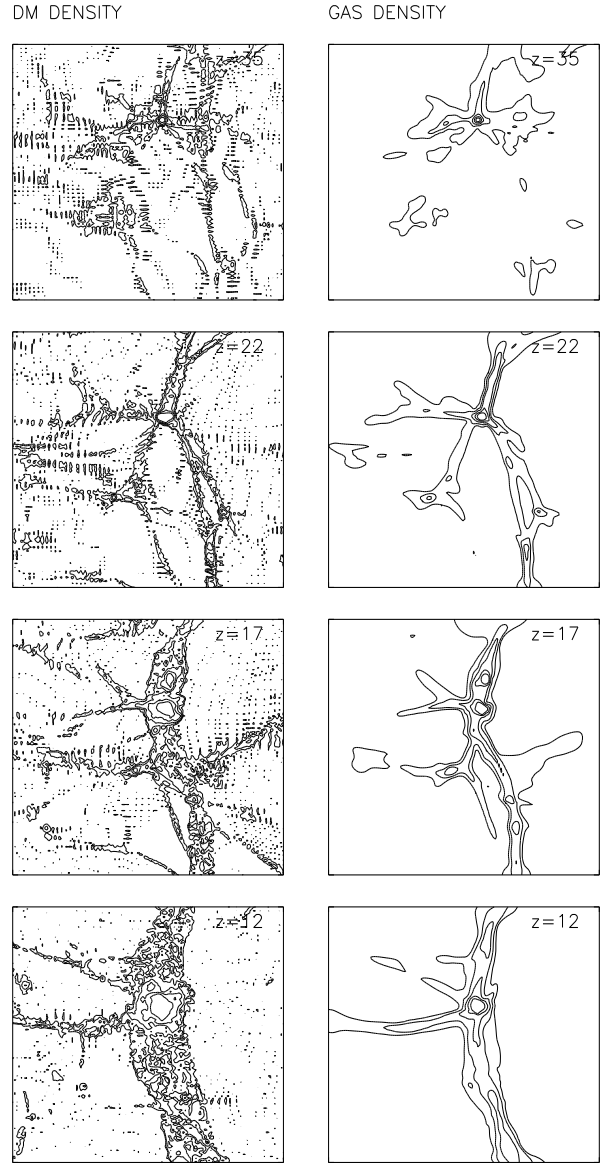,height=17cm,width=12cm}}
\caption{As in the previous figure, but for the higher resolution sub grid.
}
\label{cont_dens_sub}
\end{figure}

\begin{figure}[thb]
\centerline{\psfig{file=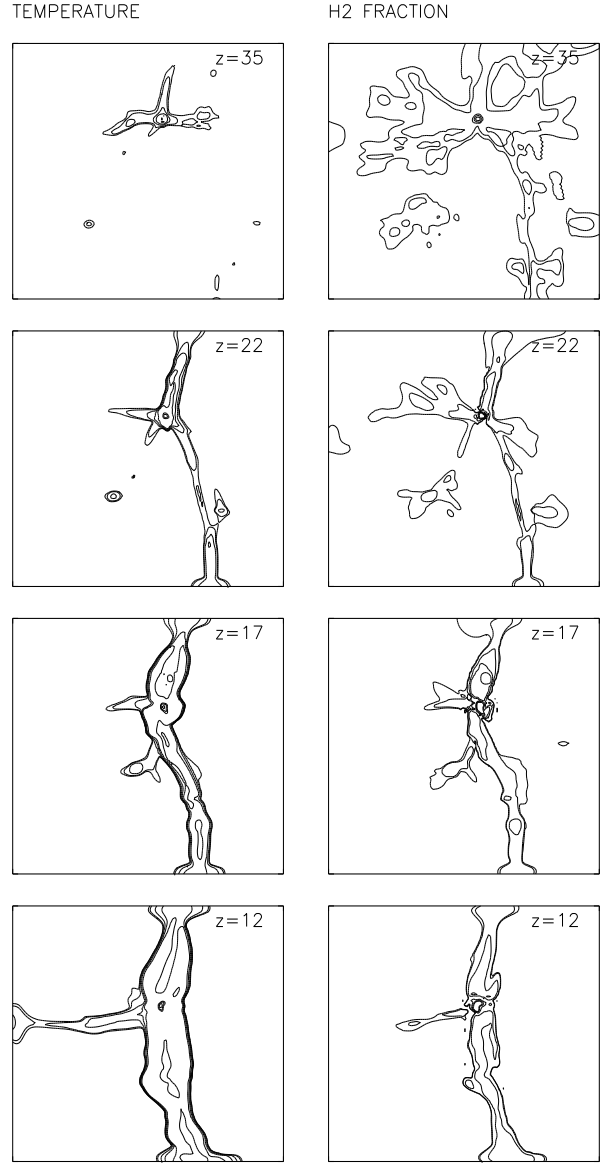,height=17cm,width=12cm}}
\caption{ Contour surfaces showing an evolution sequence for the gas
  temperature and H$_2$ fraction $n_{H2}/n_H$ at redshifts $z=$ 35,
  22, 17 and 12. Five levels are displayed for both the temperature
  ($\log T(eV)=$ -2.2, -2, -1.5, -1, -0.5) and the H$_2$ fraction
  ($\log n_{H2}/n_H =$ -5.05, -5, -4.5, -4, -3.5).  }
\label{cont_temh2_sub}
\end{figure}

Primordial stars can only form in the fraction of gas that is the able
to cool. The fraction of cooled to total baryonic mass in the
collapsed structures is only few percent.  This is illustrated in
Figure \ref{M_coolvsz} which shows the ratio of cooled mass to
$M_{200}$ as a function of redshift, where $M_{200}$ is the mass found
within a spherical structure with a mean overdensity of 200.  The
amount of cooled gas is computed by adding up all the mass within the
radius at which the temperature of the radially averaged profiles
starts to decrease.
The bigger the box  the more gas is  able to finally cool, in agreement
with the  analytical picture where higher  virial temperatures lead to
larger molecular   hydrogen abundances   and  hence   shorter  cooling
timescales.
The apparent slower growth   of  the structure in   the  $4\sigma1024$
simulation is an artifact of grid resolution since the cooling flow is
not resolved as well as in the smaller grids.

Padoan \etal (1997) put forward a theory for the origin of GCs in
which the non-equilibrium formation of \HH filters mass scales $\gsim
10^8 \Ms$ out of a CDM spectrum. In their model, these clouds convert
about 0.3\% of their total mass into a gravitationally bound stellar
system and $\sim 2\%$ into halo stars.  The remaining gas is enriched
with metals and dispersed away from the GC by supernova explosions.
Some of their assumed initial conditions are in agreement with what we
find in our simulations.  At redshift $z\sim30$ a baryonic mass of
about $10^3 \Ms$ has cooled to $T \sim 300\K$ (in the $4\sigma128$
simulation).  Although our numerical experiments fail to resolve the
hydrodynamics within the cooled component, this sub--Jeans mass cold
cloud may well host a complex supersonic random velocity field
produced by isothermal shocks which are triggered by merger events
with smaller surrounding structures.  The gas stays isothermal since
the \HH cooling time scale is short, as can be seen in
Figure~\ref{fig:evol_cool}.  Therefore, this small cloud can likely
form stars through the mechanism suggested by Padoan (1995).  A
perturbation of total mass $M$ contains $\Omega_BM$ baryons, of which
a fraction $f$ can cool. Defining the star formation efficiency as
$\epsilon$, the resulting mass in stars can be written as 
$M_s=\Omega_B f\epsilon M= 5\tento{-5} M (\Omega_B/0.05) (f/0.05)(\epsilon/0.02)$. 
If one adopts the value of Padoan~\etal~(1997) for  
the star formation efficiency ($\epsilon=0.02$) , the mass 
in stars formed in the smallest simulated structures of total mass
$6.2\tento{5}\Ms$ ($128\kpc$ box sizes) might be quite small $\sim
30\Ms$. Hence, the first stellar systems formed in hierarchical
structure formation scenarios, might be much smaller than thought
previously.  

Only two convincing detections of \HH Lyman Werner Band absorption
features in damped Ly$\alpha$ quasar absorption systems exist to date.
The one system of Ge \& Bechtold (1997) toward Q0013--004 has a very
high reported \HH fraction of $f_{H_2} = 2N(H_2)/[2N(H_2) + N(HI)] =
0.22 \pm 0.05$ arguing for very efficient \HH formation on dust grains.
The second system found earlier by Levshakov and Varshalovich (1985)
and confirmed by Foltz~\etal (1988) toward PKS~0528--250 shows an \HH
fraction of $\sim 2\tento{-3}$ which is a typical value for primordial
structure formation as we have seen above. Since the \HH column density
in this system is $\sim 10^{18}\cm^{-2}$ it will be self--shielded
from intergalactic radiation allowing the \HH formed during the initial
collapse of that system to survive the UV flux from the quasar.
Unfortunately this system shows an absorption redshift greater than the
emission redshift of the background quasar and is therefore likely to
be physically influenced by the quasar. Furthermore, its high Doppler
parameter of $b=100\kms$ found for the HI component suggests that the
absorption originates in a systems larger then the objects discussed
here.  It is nevertheless interesting that the \HH fraction, HI
column density, and the reported excitation temperature of $T_{exc}\sim
100\K$ (Foltz \etal 1988) can be found for halos with virial
velocities exceeding $10\kms$ that collapse at high redshifts, $z\gsim
10$ (see Abel \& Mo 1997). I.e.  systems as the ones simulated in this
investigation.

None of the investigations that   have implemented a  self--consistent
treatment of the  \HH chemistry and  cooling with the hydrodynamics of
collapsing structures ({\it   ie.} Anninos  \& Norman,  1996;   Haiman
\etal, 1996;  this work;   Abel   \etal  1997) have formed    baryonic
overdensities  greater than $10^5$.  For  some  of these studies, this
can  perhaps  be  attributed  to  the lack   of numerical  resolution.
However, we would like  to  stress that currently  no self--consistent
model of primordial star formation  exists. As a  next step towards an
understanding of the origin  of the first star in  the universe we are
currently implementing our chemistry/cooling model into the Berger and
Oliger  (1984) adapted mesh  refinement (AMR)  scheme to provide
dramatic improvements in  the  dynamic range  of both mass  and length
scales.   This  approach will allow   us to resolve   the core  of the
cooling structures and address questions that remain unanswered by this
and previous studies.

\acknowledgments

We happily acknowledge discussions with M.J.  Rees, Greg Bryan, Simon
White, Zoltan Haiman, and Max Tegmark, and thank Stefano Tin\'e and
Stepphen Lepp for communicating their unpublished work on \HH cooling
with us.  This work is done under the auspices of the Grand Challenge
Cosmology Consortium (GC3) and supported in part by NSF grant
ASC--9318185.  The simulations were performed on the CRAY--C90 at the
Pittsburgh Supercomputing center, and the CONVEX--3880 at the National
Center for Supercomputing Applications.

\end{document}